\documentclass[12pt]{article}
\usepackage{graphicx}
\begin{document}
\baselineskip=5.6mm
\newcommand{\be} {\begin{equation}}
\newcommand{\ee} {\end{equation}}
\newcommand{\Be} {\begin{eqnarray}}
\newcommand{\Ee} {\end{eqnarray}}
\renewcommand{\thefootnote}{\fnsymbol{footnote}}
\def\a{\alpha}
\def\b{\beta}
\def\g{\gamma}
\def\G{\Gamma}
\def\d{\delta}
\def\D{\Delta}
\def\e{\epsilon}
\def\k{\kappa}
\def\l{\lambda}
\def\L{\Lambda}
\def\t{\tau}
\def\om{\omega}
\def\Om{\Omega}
\def\s{\sigma}
\def\lg{\langle}
\def\rg{\rangle}
\noindent
\begin{center}
{\Large
{\bf
Aging effects in simple models for glassy relaxation}
}\\
\vspace{0.5cm}
\noindent
{\bf Gregor Diezemann} \\
{\it
Institut f\"ur Physikalische Chemie, Universit\"at Mainz,
Welderweg 11, 55099 Mainz, FRG
\\}
\end{center}
\vspace{1cm}
\noindent
{\it
Aging effects in the two-time correlation function and the response function after a
quench from a high temperature to some low temperature are considered for a simple
kinetic random energy model exhibiting stretched exponential relaxation.
Because the system reaches thermal equilibrium for long times after the quench, all
aging effect are of a transient nature.
In particular, the violations of the fluctuation-dissipation theorem are considered
and it is found that the relation between the response and the two-time correlation
function depends on another function, the so-called asymmetry.
This asymmetry vanishes in equilibrium but cannot be neglected in the aging regime.
It is found that plots of the integrated response versus the correlation function are
not applicable to quantify the violations of the fluctuation-dissipation theorem in
this particular model.
This fact has its origin in the absence of a scaling form of the correlation.
}

\vspace{0.5cm}
\vspace{1cm}
\section*{I. Introduction}
In the last decade, much attention has been paid to the out-of-equlibrium dynamics of
glassy systems.
In particular, the deviations from the fluctuation-dissipation theorem (FDT) have
been studied for a variety of models\cite{CR03}.
Usually, the extent of the violations of the FDT are quantified via the so-called
fluctuation-dissipation ratio (FDR) $X(t,t_w)$, defined by
\be\label{X.FDT}
R(t,t_w)={X(t,t_w)\over T}{\partial C(t,t_w)\over\partial t_w}.
\ee
Here, $R(t,t_w)$ denotes the linear response of a dynamical variable $M$ to a field
that has been applied to the system at $t_w$ and $C(t,t_w)$ is the corresponding
auto-correlation function. In equilibrium, $X(t,t_w)\!\equiv\!1$ and the FDT is
recovered. 
In non-equilibrium situations, it is tempting to use the FDR for the definition of an
effective temperature characterizing the state of the system\cite{CKP97}.
For some models of glassy relaxation, it has been found that in the scaling regime
$X(t,t_w)$ is a function of the correlation alone, $X(C)$.
Furthermore, often $X(C)\!=\!1$ holds for short times, whereas $X(C)\!<\!1$ in the
long-time sector, implying an effective temperature which is higher than the bath
temperature.
In addition to a number of analytical calculations, several molecular dynamics
simulations have been performed on model glassforming liquids, for a recent review
see ref.\cite{barrat03}.
Of course, in order to allow a meaningful definition of a thermodynamic quantity like
a temperature, $X(t,t_w)$ should be independent of the dynamical variable used for
its calculation.
In case that so-called neutral variables $M$\cite{FS02} are used, $X(t,t_w)$ often is
found to be independent of $M$ and in some cases the long time limit $X^\infty$
allows the definition of an effective temperature.

In a previous paper\cite{gregfdt05}, I considered in detail the relation between the
response and the correlation for stochastic Markov processes in which case the
dynamics is described by a master equation\cite{vkamp81}.
A treatment of FDT violations for stochastic models with Langevin dynamics has been
given earlier\cite{CKP94}.
In both cases it is found that the response is not determined by time derivatives of
the correlation alone but that an additional function $A(t,t_w)$, called asymmetry,
is needed to determine the response.
In the context of master equations, the asymmetry has been discussed for the first
time in refs.\cite{SH89, HS90} in the context of the aging dynamics in spin glasses.
It was shown in ref.\cite{gregfdt05} that different kinds of dynamical
variables can give rise to qualitatively different results regarding the
asymmetry and therefore also the FDR $X(t,t_w)$.
Two special choices for the dynamical variable $M(t)$ have been considered.
One class of variables randomizes completely with any transition among the
states of the system.
Another class of variables is chosen in such a way that there is no correlation
to the transitions among the states at all. 
The first type of variable is standard in the investigation of trap
models\cite{trap,BD95,Ritort03,soll03}.
Even though in both cases the variables can be chosen as neutral, the behavior
of the relevant dynamic quantities can be different.
In general, the asymmetry vanishes for the first class of variables under
very mild conditions but not for uncorrelated variables.
In the trap model both kinds of variables yield identical results. 
Furthermore, all dynamical quantities show a $(t/t_w)$-scaling in the long-time
limit and the FDR $X(t,t_w)$ as determined by eq.(\ref{X.FDT}) coincides with
the slope in a fluctuation-dissipation (FD) plot of the integrated response
versus the correlation function\cite{Ritort03}.

In the present paper, I will consider another class of models in which equilibrium is
reached after long times.
Therefore, all aging effects are of a transient nature but the FDT still is violated
for short enough waiting times.
It will be shown that in a simple model the asymmetry does not vanish and has a 
strong impact on the FDR in the aging regime.
Additionally, it turns out that the FDR calculated directly from its definition
deviates from the slope in FD plots.
After a short review of the calculation of the response and the correlation for
Markov processes I will discuss the random energy model\cite{derrida80} with a
kinetic rule that has been considered earlier by Koper and Hilhorst\cite{KH89} for 
the relaxation in equilibrium.
\section*{II. FDT violations for stationary Markov processes}
In this chapter I briefly recall the derivation of the fluctuation-dissipation
relation for stationary Markov processes as it has been derived in
ref.\cite{gregfdt05}.
I will concentrate on stationary Markov processes for simplicity, but also
nonstationary processes can be treated in the same way.

In order to calculate the fluctuation-dissipation relation both, the response and the
correlation are needed.
In all following calculations it is assumed that the system is quenched from a high
temperature to a low 'working' temperature in the beginning of the experimental
protocol. After a waiting time $t_w$ has evolved after the quench, either a field
$H\d(t-t_w)$ is applied and the response is measured or the auto-correlation function
is monitored.
For a dynamical variable $M(t)$ the correlation function is given by:
\be\label{C.t.tw}
C(t,t_w)=\lg M(t)M(t_w)\rg=\sum_{k,l}M_kM_lG_{kl}(t-t_w)p_l(t_w)
\ee
Here, $G_{kl}(t-t_w)$ is the conditional probability to find the system in state 'k'
at time $t$, provided it was in state 'l' at time $t_w$.
The population $p_l(t_w)=\sum_nG_{ln}(t_w)p_n^0$ gives the probability to find the
system in state 'l' at the beginning of the measurement.
The initial high-temperature state of the system is assumed to be described by a 
fixed set of populations, $p_k^0\!=\!p_k(t\!=\!0)$ with $\sum_kp_k^0\!=\!1$.
Furthermore, $M_k$ denotes the value of $M$ in state 'k'.
In the above expressions, I used a discrete notation and the term 'state' is meant to
represent the stochastic variable under consideration.
In many models for glassy relaxation, this variable will be a (free) energy.
The conditional probability $G_{kl}(t-t_w)$ obeys a master equation\cite{vkamp81}:
\be\label{ME}
{\partial\over\partial t}G_{kl}(t)=
-\sum_nW_{nk}G_{kl}(t)+\sum_nW_{kn}G_{nl}(t)
\ee
with $W_{kl}$ denoting the rates for a transition from state $l$ to state $k$.

For a calculation of the linear response to a field $H(t)=H\d(t-t_w)$ conjugate to
$M$,
\be\label{R.def}
R(t,t_w)=\left.{\d\lg M(t)\rg\over\d H(t_w)}\right|_{H=0}
\ee
one has to fix the dependence of the transition rates on the field $H$.
There is no general way to do this and here, as in ref.\cite{gregfdt05}, I use the
following form of multiplicatively perturbed transition probabilities\cite{Ritort03}:
\be\label{kap.h}
W^{(H)}_{kl}=W_{kl}e^{\b H \left(\g M_k-\mu M_l\right)}
\ee
Here, $\g$ and $\mu$ are arbitrary parameters.
If $\mu+\g\!=\!1$ holds additionally, then the rates $W^{(H)}_{kl}$ obey detailed
balance also in the presence of the field.

Without going into the details of the calculation of the response according to
eq.(\ref{R.def}), I only mention that time-dependent perturbation theory is used to
calculate $G^{(H)}_{kl}(t)$ in linear order with respect to the
field\cite{FS90}.
In case of stationary Markov processes described by a continuous time master equation
the result can be cast into the form\cite{gregfdt05}:
\be\label{FDT.gen}
R(t,t_w)=\b\left[\g{\partial C(t,t_w)\over\partial t_w}
		 -\mu{\partial C(t,t_w)\over\partial t}
		 -\g A(t,t_w)\right]
\ee
with the asymmetry
\be\label{A.def}
A(t,t_w)=\sum_{k,l,n}M_kM_lG_{kn}(t-t_w)
		     \left[W_{ln}p_n(t_w)-W_{nl}p_l(t_w)\right]
\ee
It is important to point out that the asymmetry $A(t,t_w)$ cannot be related to a
time derivative of the correlation function and therefore the response is not
determined by $C(t,t_w)$ alone in the general case.
If the system is in equilibrium, one has $A_{eq}(t,t_w)\!\equiv\!0$.
This can be achieved in two ways. Either the system has been prepared in an
equilibrium state initially, i.e. $p_k^0\!=\!p_k^{eq}$, or the system has reached
equilibrium after a sufficiently long waiting time.
In both cases one has $p_l(t_w)\!=\!p_l^{eq}$ and use of the detailed balance
condition $W_{ln}p_n^{eq}=W_{nl}p_l^{eq}$ shows that $A_{eq}(t,t_w)\!=\!0$.
In these situations $C(t,t_w)$ and $R(t,t_w)$ are time-translational invariant and
one finds
\be\label{fdt.eq}
R_{eq}(t)=-\b(\g+\mu){d C_{eq}(t)\over d t}
\ee
which for $\mu\!=\!1-\g$ is just the well known FDT.

In the following, I will consider the same types of dynamical variables that have
been discussed in detail in ref.\cite{gregfdt05}.
One class has been denoted as 'uncorrelated' and the other one as 'randomizing'.
The meaning of these terms is the following.
An uncorrelated variable is completely decoupled from the transitions among the
states of the system. This is the choice made by Koper and Hilhorst\cite{KH89} in
their treatment of a kinetic random energy model.
In contrast, a variable for which every transition among the states of the system
gives rise to a complete randomization is called a randomizing variable.
This is the type of variable that is used in most calculations for trap models.
In the following, I will restrict the discussion to variables the values of which are
independent of the state of the system, $M_k\!=\!M$ $\forall k$, and furthermore
assume distributions of zero mean and unit variance,
\be\label{M.mit}
\lg M\rg=0\quad\mbox{and}\quad \lg M^2\rg=1
\ee
With this choice one explicitly finds for the two types of variables\cite{gregfdt05}
\Be\label{C.A.unkorr}
\mbox{uncorrelated variables:}\quad
C(t,t_w)=&&\hspace{-0.6cm}\sum_kG_{kk}(t-t_w)p_k(t_w)\nonumber\\
A(t,t_w)=&&\hspace{-0.6cm}\sum_{k,l}G_{kl}(t-t_w)
                   \left[W_{kl}p_l(t_w)-W_{lk}p_k(t_w)\right]
\Ee
and
\Be\label{C.A.random}
\mbox{randomizing variables:}\quad
\Pi(t,t_w)=&&\hspace{-0.6cm}\sum_ke^{-\k_k(t-t_w)}p_k(t_w)
\quad\mbox{with}\quad\k_k=\sum_lW_{lk}\nonumber\\
A(t,t_w)=&&\hspace{-0.6cm}0
\Ee
In the latter case, I denoted the correlation function by $\Pi(t,t_w)$, because this 
function is identical to the probability that the system has not left the state 
occupied at $t_w$ considered in the classical treatment of trap 
models\cite{trap,MB96}. 
For trap models one has $G_{kk}(t-t_w)=e^{-\k_k(t-t_w)}+{\cal O}(N^{-1})$ and
therefore uncorrelated variables yield the same result as randomizing variables
for large $N$.

Before closing this section, some comments are in order. 
One comment concerns the values of the parameters $\g$ and $\mu$. 
Even if one enforces detailed balance to hold in the presence of the field,
i.e. if $\g+\mu=1$ holds, there is no general way to fix them. 
Only in some special cases one has some arguments based on the physics of the
corresponding model. 
For instance, if one considers particles hopping on a (disordered) lattice, one
would naturally choose $\g=\mu=1/2$. 
Slightly more generally, one would expect the choice $\g=\mu=1/2$ to be
meaningful if the master equation considered has a form allowing a Kramers-Moyal
expansion\cite{vkamp81}. 
These arguments, however, are not sufficient to allow the determination of $\g$
and $\mu$ in general. 

Regarding the asymmetry $A(t,t_w)$, it was stated above that this quantity
cannot be related to a time derivative of the correlation function. 
Furthermore, there is no obvious physical interpretation of $A(t,t_w)$. 
Apparently, it has to be related to some correlations among the dynamical
variables as otherwise it would not vanish in case of randomizing variables. 
However, $A(t,t_w)$ appears quite strongly to depend on both, the model
considered and the choice of variables. 
At present, it is not clear whether it is related to some other dynamical
function of general importance.
\section*{III. A kinetic random energy model}
In the following, I will apply the results of the preceeding section to a very simple
model which allows to calculate all quantities analytically.
The model exhibits interrupted aging only, reaching equilibrium in the long-time
limit.
Such kind of models are interesting to study because in canonical glasses the
situation is somewhat similar.
If a glass-forming liquid is quenched to a temperature below the glass transition
temperature, this means that the re-equilibration time exceeds typical experimental 
time scales.
However, one expects the system to reach (metastable) equilibrium for very long
times.
A similar situation may appear in computer simulation studies of model
glass-formers.
Here, the available computer time sets the time scale for following the equilibrium
properties of the system.

I will treat a special case of a kinetic random energy model\cite{derrida80},
following the treatment of Koper and Hilhorst\cite{KH89}, who considered various
forms of the transition rates.
For the purpose of the present discussion the simplest choice
\be\label{k.kl.rcm}
W_{kl}=\k_0B_k=\k_0e^{-\b\e_k}
\ee
is sufficient, where $\k_0$ is a rate constant, to be set to unity in the
following, $\b\!=\!T^{-1}$, and $B_k$ is a Boltzmann factor.
According to eq.(\ref{k.kl.rcm}) the $W_{kl}$ only depend on the destination
state
of the transition and the corresponding ME can easily be solved with the result
\be
G_{ik}(t)\!=\!Z(\b)^{-1}B_i+\left[\d_{ik}-Z(\b)^{-1}B_i\right]\exp{(-Z(\b)t)}
\ee
with the partition function $Z(\b)\!=\!\sum_k B_k$.
The equilibrium populations are given by
$p_i^{eq}\!=\!G_{ik}(t\!\to\!\infty)\!=\!Z(\b)^{-1}B_i$.
Therefore, the system reaches equilibrium for long times and all aging
phenomena are of a transient nature.

As already noted, I will use a distribution of 'magnetizations' $M$ with
$\lg M\rg\!=\!0$ and $\lg M^2\rg\!=\!1$ throughout the calculations.
In addition, the initial populations $p_i^0$ will be chosen as $p_i^0\!=\!N^{-1}$,
appropriate for a quench from high temperatures in the beginning of the experimental
protocol.
The final temperature is chosen to be lower than the phase transition temperature
$T_c$ of the random energy model\cite{derrida80}.
The calculation of the quantities of interest proceeds in the following way.
First one calculates the correlation and asymmetry according to
eqns.(\ref{C.A.unkorr},\ref{C.A.random}) using the expression given above for the
Greens function $G_{kl}(t)$ and $p_k(t_w)=\sum_lG_{kl}(t_w)p_l^0$.
The quantities calculated this way depend on the Boltzmann factors, which will be
denoted by a superscript $B$, e.g. $C^{(B)}(t,t_w)$ etc..
These expressions have to be averaged over the distribution of Boltzmann factors, 
which derive from the distribution of the random energies.
Below $T_c$ the random energies are exponentially distributed and therefore the
corresponding distribution of Boltzmann factors $B_k$ is given by
$p(B)=(v/N)\times B^{-1-x}$ for $(v/N)^{(1/x)}\!<\!B\!<\!\infty$ and
$p(B)\!=\!0$ otherwise\cite{KH89}.
Here, $x\!=\!T/T_c$, $v\!=\!(2\sqrt{\pi\log{2}})^{-1}$ and $N$ denotes the
number of random energies, to be sent to infinity at the end of a calculation.
In a shorthand notation one then has
$F(t,t_w)=\lg F^{(B)}(t,t_w)\rg=\int\!dBp(B)F^{(B)}(t,t_w)$.
In the following, I will discuss the two classes of variables introduced above
separately.
\subsubsection*{Uncorrelated variables}
For this case one finds, using the abbreviation ${\cal Z}(\b)\!=\!Z(2\b)/Z(\b)^2$:
\Be\label{C.rem.unkorr.ungem}
C^{(B)}(t,t_w)=&&\hspace{-0.6cm}
{\cal Z}(\b) +\left[1-{\cal Z}(\b)\right]\exp{(-Z(\b)(t-t_w))}\nonumber\\
&&\hspace{-0.6cm}
+{\cal Z}(\b)\left[\exp{(-Z(\b)t)}-\exp{(-Z(\b)t_w)}\right]
\Ee
and
\be\label{A.rem.unkorr.ungem}
A^{(B)}(t,t_w)=Z(\b){\cal Z}(\b)\left[\exp{(-Z(\b)t_w)}-\exp{(-Z(\b)t)}\right]
\ee
This expression can be cast into the form
$A^{(B)}(t,t_w)=\partial_t C^{(B)}(t,t_w)+\partial_{t_w}C^{(B)}(t,t_w)$, which
allows to write the expression for the response in the form:
\be\label{R.rem.unkorr.ungem}
R^{(B)}(t,t_w)=-\b\left(\g+\mu\right){\partial C^{(B)}(t,t_w)\over\partial t}
\ee
Therefore, in this case the response depends on the parameters $\g$ and $\mu$
only via their sum. As already noted, one has $\g+\mu=1$ if it is assumed that
detailed balance holds in the presence of the field.  
In order to perform the averages of the quantities given in
eqns.(\ref{C.rem.unkorr.ungem},\ref{A.rem.unkorr.ungem},\ref{R.rem.unkorr.ungem}),
the only integrals required are $\lg e^{-Z(\b)t}\rg$ and
$\lg{\cal Z}(\b)e^{-Z(\b)t}\rg$.
The calculation of these integrals has been performed in ref.\cite{KH89} with
the result:
\Be\label{Phi.Psi.def}
&&\Phi(t)=\lg e^{-Z(\b)t}\rg=\exp{(-\tilde{v}t^x)}\nonumber\\
&&\Psi(t)=\lg{\cal Z}(\b)e^{-Z(\b)t}\rg
=(1-x)\left[\Phi(t)-\tilde{v}^{1/x}t\G(1-1/x;\tilde{v}t^x)\right]
\Ee
where $\tilde{v}\!=\!v\G(1-x)$ with the Gamma function $\G(a)$.
Furthermore, $\G(a;b)$ denotes the incomplete Gamma function.
In terms of these functions the averaged quantities read:
\Be\label{C.R.unkorr}
C(t,t_w)=&&\hspace{-0.6cm}
1-x+\Phi(t-t_w)-\Psi(t-t_w)+\Psi(t)-\Psi(t_w)\nonumber\\
A(t,t_w)=&&\hspace{-0.6cm}
\partial_t\Psi(t)-\partial_{t_w}\Psi(t_w)\\
R(t,t_w)=&&\hspace{-0.6cm}
-\b\left(\g+\mu\right)\partial_t C(t,t_w)\nonumber
\Ee
As stated above, the system reaches equilibrium for long waiting times
$t_w\gg\t_{\rm eq}$, where $\t_{\rm eq}$ denotes the re-equilibration time.
In this case, one finds from eq.(\ref{C.R.unkorr}), assuming $\t\!=\!(t-t_w)$ to be
finite:
\be\label{C.R.unkorr.eq}
C_{\rm eq}(\t)=1-x+\Phi(\t)-\Psi(\t)
\quad\mbox{and}\quad
R_{\rm eq}(\t)=-\b\left(\g+\mu\right)
                   {\partial C_{\rm eq}(\t)\over\partial\t}
\ee
and $A_{\rm eq}(t,t_w)\!\equiv\!0$, as expected.
Note that $C(t,t_w)$ can be cast into the form
$C(t,t_w)\!=\!C_{\rm eq}(t-t_w)+C_{\rm ag}(t,t_w)$.
However, $C_{\rm ag}(t,t_w)$ does not exhibit a $(t/t_w)$-scaling.

The behavior of $C_{\rm eq}(\t)$ has been discussed in detail by Koper and
Hilhorst\cite{KH89}.
$C_{\rm eq}(\t)$ decays in a stretched exponential fashion,
$C_{\rm eq}(\t)\!=\!C_{\rm eq}^\infty+
(1-C_{\rm eq}^\infty)\exp{\{-(t/\t_{\rm eq})^{\b_{\rm eq}}}\}$ with
$C_{\rm eq}^\infty\!=\!(1-x)$.
Also for finite $t_w$ $C(t,t_w)$ can be fitted to a Kohlrausch function
$C(t_w+\t,t_w)\!=\!C^\infty(t_w)+(1-C^\infty(t_w))\exp{\{-(\t/\t_K)^{\b_K}}\}$ where
the plateau-value $C^\infty(t_w)$ equals zero for vanishing $t_w$ and smoothly
approaches $C_{\rm eq}^\infty$ for long waiting times.
This behavior is shown in Fig.1a, where $C(t_w+\t,t_w)$ is plotted versus $\t$ for
various waiting times (upper panel).
The dotted lines represent least-square fits to the mentioned Kohlrausch function.
Only for intermediate $t_w$ there are some minor deviations visible.
In the lower panel of Fig.1a the correlation function is shown in a normalized way so
that it decays from unity to zero.
It is evident from that plot that the relaxation time for intermediate waiting times
exceeds the equilibrium relaxation time.
This fact is further substantiated in Fig.1b, where the results of Kohlrausch fits to
$C(t_w+\t,t_w)$ are shown.
The relaxation time $\t_K(t_w)$ first increases and then decreases again before it
tends to the limit $\t_{eq}$.
The stretching parameter also first increases and goes through a maximum before it
reaches its equilibrium value.
This latter fact means that the distribution of populations first narrows and then
broadens again as a function of the waiting time.
A qualitatively similar behavior has also been found in some computer
simulations\cite{SVS04} and in a free-energy landscape model for glassy
relaxation\cite{Elmaging}.
Another important feature that becomes evident from Fig.1 is that the correlation
function does not exhibit a $(t/t_w)$-scaling, cf. the lower panel of Fig.1a.
For other temperatures the behavior is the same.

Next, the asymmetry is considered.
According to eq.(\ref{C.R.unkorr}), $A(t,t_w)$ reaches a plateau determined by
$[-\partial_{t_w}\Psi(t_w)]$ for long times $t$ and finite $t_w$.
Furthermore it can be shown, that $A(t_w+\t,t_w)\sim\t$ for small $\t$.
The asymmetry $A(t_w+\t,t_w)$ is plotted versus $t$ in Fig.2 for various values of 
$t_w$ for $x\!=\!0.3$. 
The behavior for other values of $x$ is very similar.
The linear behavior at short times is seen in the lower panel, where a logarithmic 
scale is used. 
It is evident that the cross-over from the linear behavior to the plateau takes place 
around $t_w$, $\t\sim t_w$. 
It is obvious from Fig.2 that for uncorrelated variables the asymmetry significantly 
contributes to the response for short waiting times and thus cannot be neglected. 

According to eqns.(\ref{X.FDT}) and (\ref{C.A.unkorr}) the FDR in the present case is 
given by: 
\be\label{X.rem.unkorr}
[\g+\mu]^{-1}X(t,t_w)=-{\partial_tC(t,t_w)\over\partial_{t_w}C(t,t_w)}
=1+{\partial_t\Psi(t)-\partial_{t_w}\Psi(t_w)\over
\partial_t\Phi(t-t_w)-\partial_t\Psi(t-t_w)+\partial_{t_w}\Psi(t_w)}
\ee
This FDR may be considered as a function of either $\t\!=\!(t-t_w)$ or as a function 
of $t_w$. 
In the first case, one finds the following limits ($\mu\!=\!1-\g$): 
\be\label{X.tau.limits}
X(t_w+\t,t_w)\to 1\ ;\ \t\ll t_w\quad\mbox{and}\quad
X(t_w+\t,t_w)\to 0\ ;\ \t\gg t_w
\ee
Similarly, if $X(t,t_w)$ is considered as a function of $t_w$, as has been suggested 
in ref.\cite{FS02}, one finds 
\be\label{X.tw.limits}
X(t,t_w)\to 1\ ;\ t_w\to t\quad\mbox{and}\quad
X(t,t_w)\to 0\ ;\ t_w\to 0
\ee
Furthermore, one finds from the limiting behavior of the function $\Psi(t)$ that
$X(t,t_w)\!\sim\!t_w^{1-x}$ for small $t_w$.

This behavior of the FDR is illustrated in Fig.3 for $x\!=\!0.3$.
In the upper panel $X(t_w+\t,t_w)$ is plotted versus $\t$ for several values of the
waiting time $t_w$.
It is seen that the crossover from $X\!=\!1$ to $X\!=\!0$ takes place around
$\t\!\sim\!t_w$.
This means that for finite $t_w$ one has $X^\infty(t_w)\!=\!0$.
The lower panel of Fig.3 shows $X(t,t_w)$ as a function of $t_w$ for several values
of the later time $t$.
It is evident how $X$ approaches the equilibrium value $X\!=\!1$ for $t_w\!\to\!t$.
The dotted line shows the mentioned $t_w^{1-x}$-behavior for $t_w\!\ll\!t$.

For the integrated response $\chi(t,t_w)\!=\!\int_{t_w}^t\!dt'R(t,t')$ one finds:
\be\label{Chi.t.tw}
\chi(t,t_w)=\b(\g+\mu)\left[x-\Phi(t-t_w)+\Psi(t-t_w)
             -(t-t_w)\partial_t\Psi(t)\right]
\ee
where the first three terms correspond to the equilibrium response,
$\chi_{\rm eq}(\t)=\b\left[1-C_{\rm eq}(\t)\right]$ (because of 
$C_{\rm eq}(0)\!=\!1$), which is reached for $t_w\!\gg\!\t_{\rm eq}$. 
The limiting value for long times is thus given by 
$\chi_{\rm eq}(\infty)\!=\!\b x\!\equiv\!T_c^{-1}$.
For finite $t_w$, the limiting behavior of the integrated response is given by 
$\chi(t_w+\t,t_w)\!\to\!(\g+\mu)T_c^{-1}$ for $\t\!\to\!\infty$, which for 
$\mu\!=\!1-\g$ coincides with the equilibrium value. 

In Fig.4, typical FD-plots, $\b^{-1}\chi(t_w+\t,t_w)$ versus $C(t_w+\t,t_w)$, for 
$x\!=\!0.3$ (upper panel) and $x\!=\!0.6$ (lower panel) are shown. 
Here, $\t$ is used as the curve parameter and the different curves are for various 
$t_w$-values. 
Note that $C(t,t)\!=\!1$ always holds in this model. 
Furthermore, it is assumed that $\mu\!=\!1-\g$ (otherwise there would be an extra 
factor $\g+\mu$). 
The dotted lines represent slopes of ($-1$) (FDT) and of ($-x$).
From the FD-plots, for $t_w\!<\!\t_{\rm eq}$ ($\t_{\rm eq}(x\!=\!0.3)\!\simeq\!80$ 
and $\t_{\rm eq}(x\!=\!0.6)\!\simeq\!2.5$), one would naively extract a 
limiting FDR $X^\infty$ that coincides with the transition temperature $T_c$. 
For longer waiting waiting times, the initial slope of ($-1$) persists for longer 
times and the FDT is recovered. 

From the discussion of the FD-plot and the FDR it is obvious that the slope in a  
FD-plot is not related to the FDR in this simple model. 
This fact may be of importance if one cannot perform experiments over a long enough 
time scale in order to monitor the re-equilibration of the system. 
Note that one might argue that the FD-plots built from a correlation function with a 
finite long-time plateau value may give rise to erroneous results. 
However, in the present case one has $C^\infty(t_w)\!\simeq\!0$ for small $t_w$, cf. 
Fig.1a.  
\subsubsection*{Randomizing variables}
For this choice of variables one finds from eqn.(\ref{C.A.random}) after averaging 
over the random energies: 
\be\label{Pi.rcm}
\Pi(t,t_w)=\Phi(t-t_w)
\ee
where $\Phi(t)$ is given in eq.(\ref{Phi.Psi.def}). 
This means that the correlation and therefore also the response are 
time-translational invariant and for $\g+\mu\!=\!1$ the FDT holds because of 
$A_{\rm rand}(t,t_w)\!\equiv\!0$, cf. eq.(\ref{C.A.random}). 
Thus, randomizing variables do not at all allow to probe the out-of-equilibrium 
situation imposed by the quench from high temperatures into the low temperature 
phase for this simple model. 
This can be understood from the fact that the dynamics is not thermally activated for 
the simple choice (\ref{k.kl.rcm}) of the transition rates. 
Therefore, the variable has lost memory of the initial non-equilibrium situation  
completely after a single transition. 
This is in contrast to the situation in the trap model, where aging is due to the
fact that a equilibrium distribution of populations does not exist in the 
low-temperature phase\cite{MB96}.
\section*{V. Conclusions}
In the present paper I have recalled the results obtained earlier for the relation
between the correlation function and the linear response of a system exhibiting
stochastic dynamics described by a stationary continuous time Markov process.
For the calculation of the response the transition rates of the master equation are
assumed to be perturbed multiplicatively by the applied field.
In general, the response is not determined by time-derivatives of the correlation
alone, but the asymmetry $A(t,t_w)$ has to be considered additionally, which in
general cannot be related to time derivatives of the correlation function. 
This function in general only vanishes under equilibrium conditions, in which case
the FDT is recovered, if the dependence on the field of the transition rates is
chosen in a symmetric way ($\g+\mu\!=\!1$).

I considered two different classes of variables, namely variables that are
uncorrelated with the states of the system and variables which randomize completely
with each transitions among the states.
Both types of variables are neutral in the sense that their values are not correlated
with the state of the system, in particular the energy\cite{FS02}.

As an example for a simple model of glassy relaxation I considered a kinetic REM
with
an extremely simple choice for the transition rates.
This model shows intermittant aging behavior and reaches equilibrium for long times.
If uncorrelated variables are chosen, the asymmetry is finite for short waiting
times.
Furthermore, in this case the limiting slope in an FD-plot is determined by the
transition temperature of the model.
However, this slope does not coincide with the value of the FDR as determined from
its definition.
Therefore, in this case a FD-plot does not yield the correct value for the FDR.
Of course, the reason for this finding lies in the fact that the correlation does not
obey a scaling-law but rather decays in a stretched exponential manner.
Even more relevant to experimental determinations of the FDR in glassy systems may be
the fact that the results for different variables are different.
If randomizing variables are used instead of uncorrelated variables, the asymmetry
vanishes and all quantities are time-translational invariant.
Thus, one always has a FDR that equals unity in this case.

To conclude, I have shown that for some general class of models the fluctuation-
dissipation relation is determined by time-derivatives of the correlation
function and an additional function, the asymmetry $A(t,t_w)$.
In general, the asymmetry can be shown to vanish for randomizing variables under the
conditions considered usually.
For the model considered in this paper, $A(t,t_w)$ has a strong impact on the
behavior of the FDR for uncorrelated variables.
Additionally, it appears that one must be careful in the determination of the FDR
$X(t,t_w)$ from FD-plots in some cases.
\vspace{0.2cm}
\subsection*{Acknowledgements:}
I thank Uli H\"aberle for fruitful discussion and a careful reading of the 
manuscript. 
This work was supported by the DFG under Contract No. Di693/1-2.
\newpage
\newpage
\newpage
\section*{Figure captions}
\begin{description}
\item[Fig.1 : ]
{\bf a)} The correlation function for an uncorrelated variable in the kinetic REM
plotted as a function of time for $x\!=\!T/T_c\!=\!0.3$ and various values of the
waiting time $t_w$.
Upper panel: $C(t_w+\t,t_w)$ for $\log_{10}(t_w)=-5,-2,0,1,2,5$.
The dotted lines represent fits to a Kohlrausch function.
Lower panel: $C(t_w+\t,t_w)$ normalized in a way that it decays from unity to zero.
This plot shows that for intermediate waiting times the relaxation time goes through
a maximum.\\
{\bf b)} Fitting parameters $\t_K(t_w)$ (upper panel) and $\b_K(t_w)$ (lower panel)
resulting from Kohlrausch-fits
$C(t_w+\t,t_w)\!=\!C^\infty(t_w)+(1-C^\infty(t_w))\exp{\{-(\t/\t_K)^{\b_K}}\}$.
\item[Fig.2 : ]
$A(t_w+\t,t_w)$ versus $\t$, for $x\!=\!0.3$ for various waiting times
$t_w$ for the kinetic REM with an uncorrelated variable.
Upper panel: linear scale, lower panel: logarithmic scale.
In the lower panel, the dotted line is proportional to $\t$.
\item[Fig.3 : ]
Upper panel: $X(t_w+\t,t_w)$ versus $\t$ for $x\!=\!0.3$ and
$\log_{10}(t_w)=-5,-2,0,2,5$ for the kinetic REM with an uncorrelated variable.
The dotted vertical lines show the corresponding values of $t_w$.
Lower panel: $X(t,t_w)$ versus $t_w$ for $x\!=\!0.3$ and
$\log_{10}(t)=-5,-3,-1,1,3,4,5$.
The dotted line is $\propto t_w^{1-x}$.
\item[Fig.4 : ]
$\b^{-1}\chi(t_w+\t,t_w)$ versus $C(t_w+\t,t_w)$, for $x\!=\!0.3$ (upper panel) and
$x\!=\!0.6$ (lower panel) for various waiting times $t_w$ for the kinetic REM with an
uncorrelated variable. The waiting times chosen are $\log_{10}(t_w)=-5,-1,0,1,2$ for
$x\!=\!0.3$ and $\log_{10}(t_w)=-5,-1,0,2,$ for $x\!=\!0.6$.
The dotted lines represent the slopes expected for the FDT (slope: $-1$) and a
slope of $-x$.
\end{description}
\newpage
\begin{figure}
\includegraphics[width=15cm]{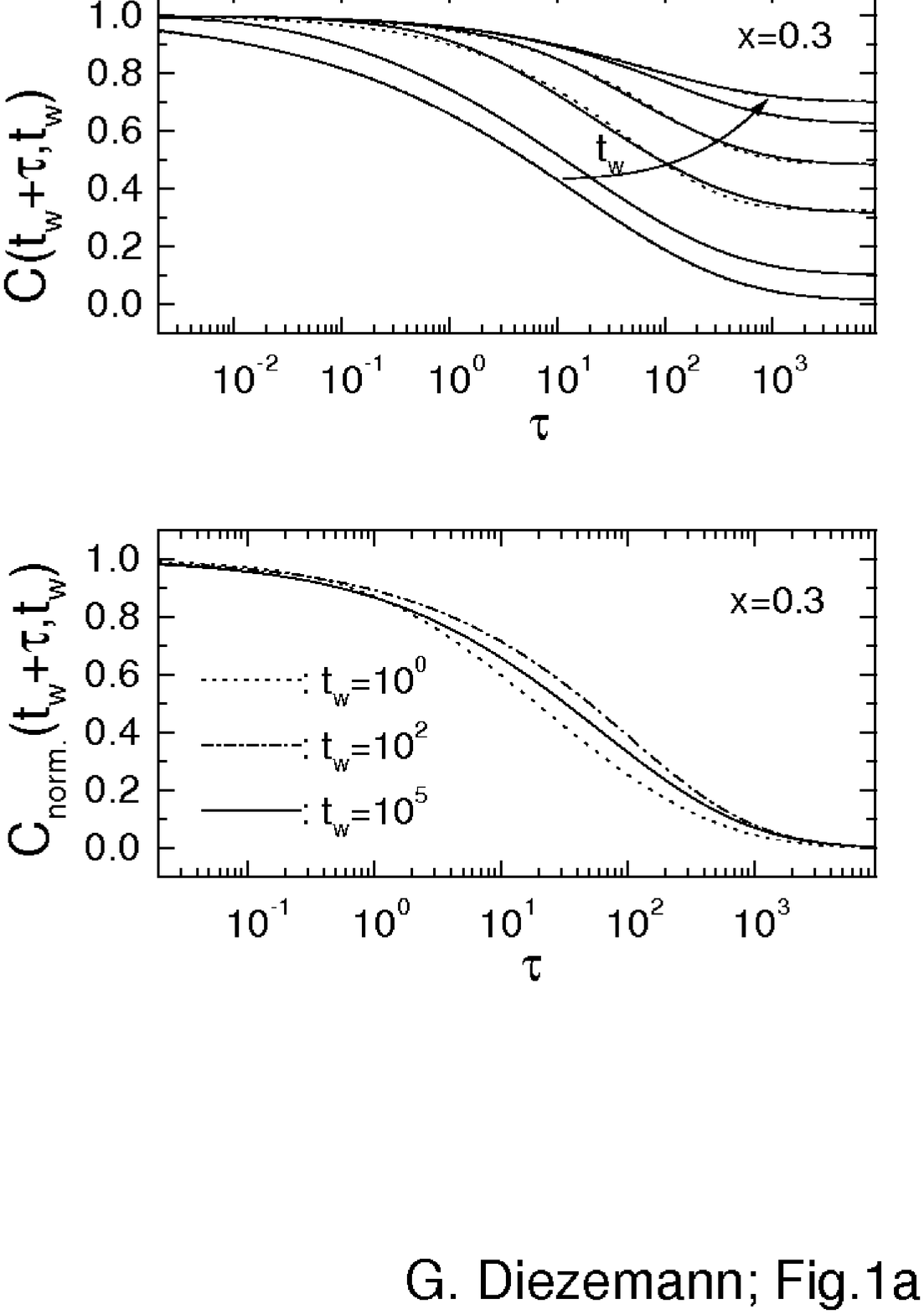}
\end{figure}
\newpage
\begin{figure}
\includegraphics[width=15cm]{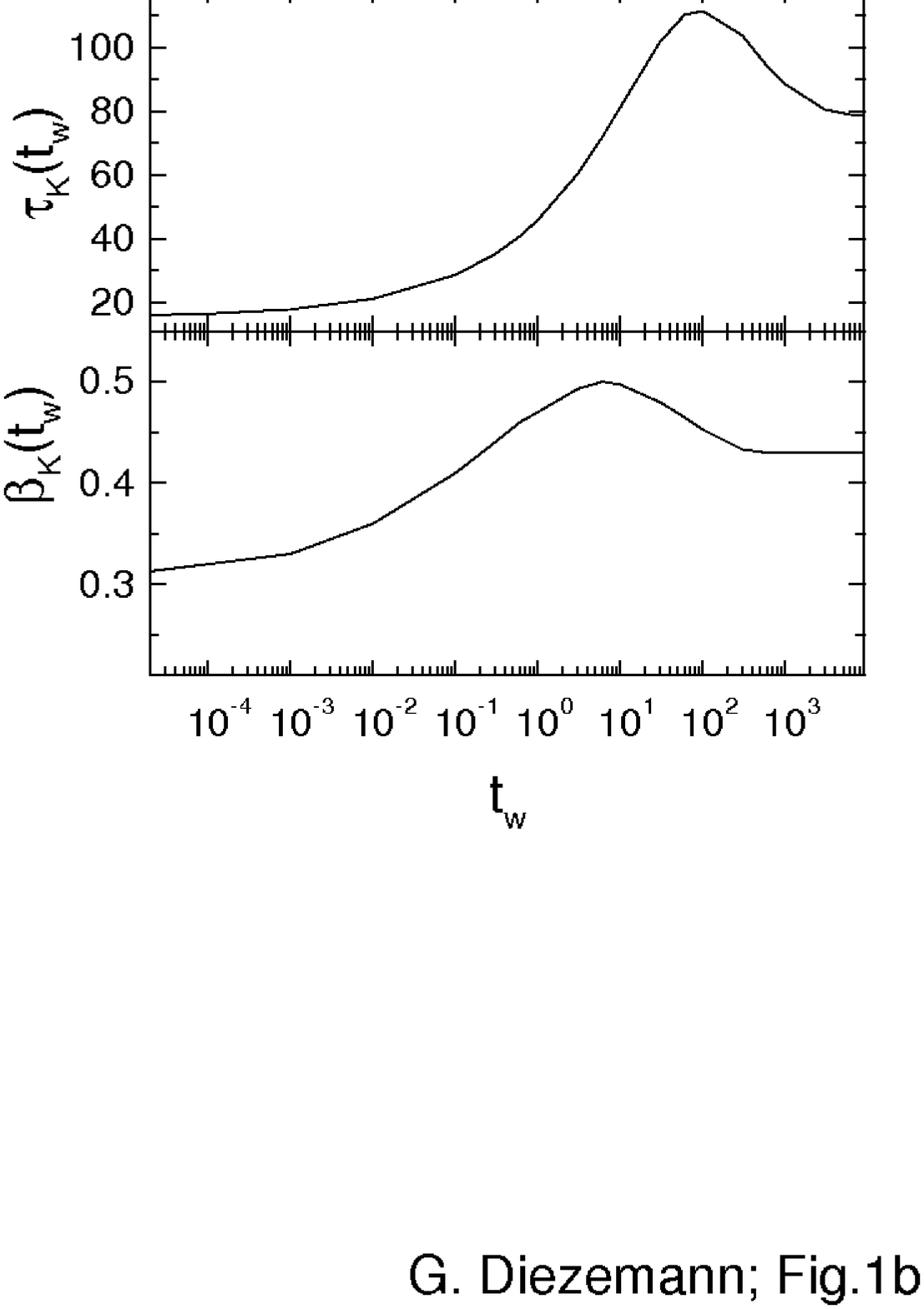}
\end{figure}
\newpage
\begin{figure}
\includegraphics[width=15cm]{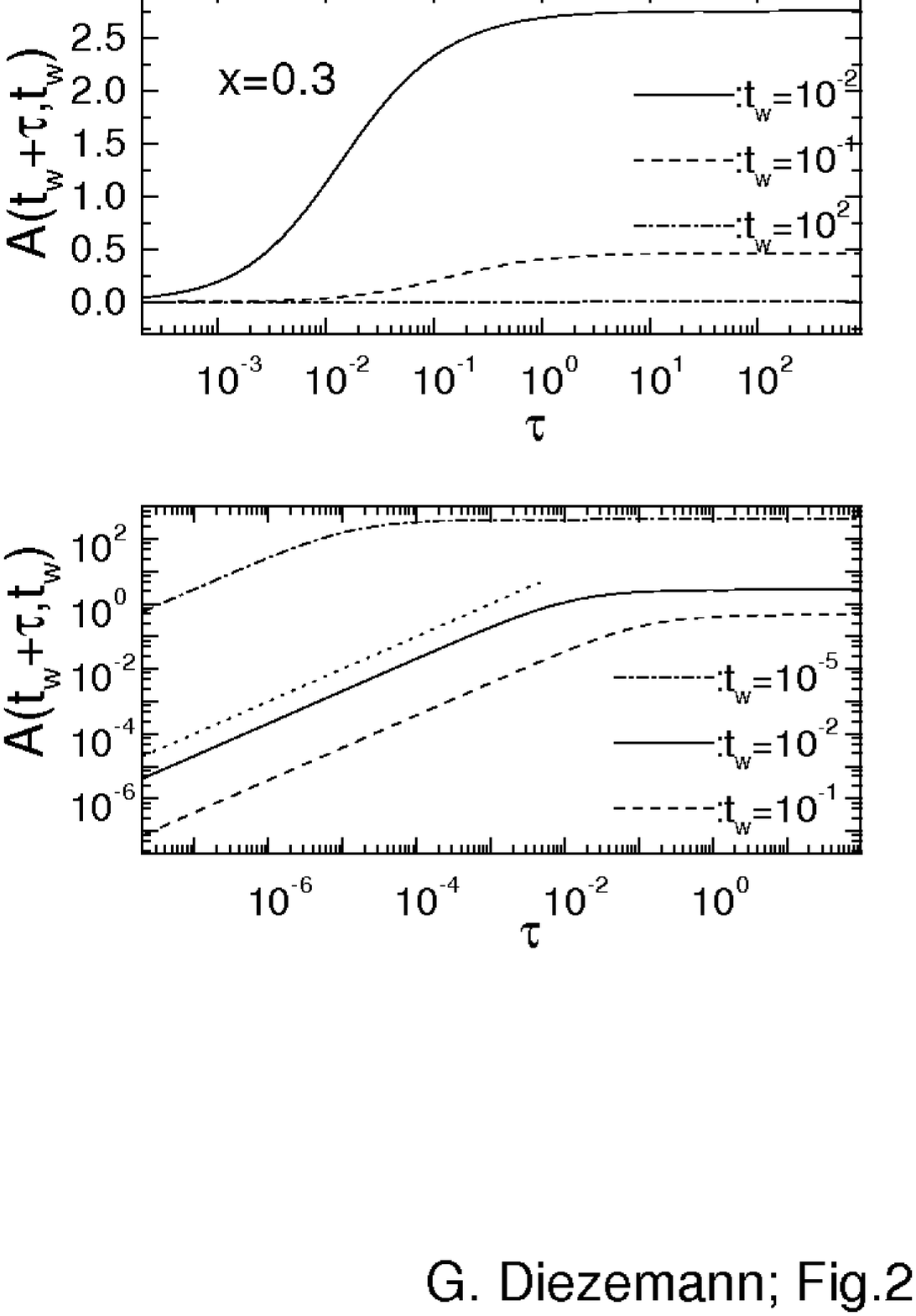}
\end{figure}
\newpage
\begin{figure}
\includegraphics[width=15cm]{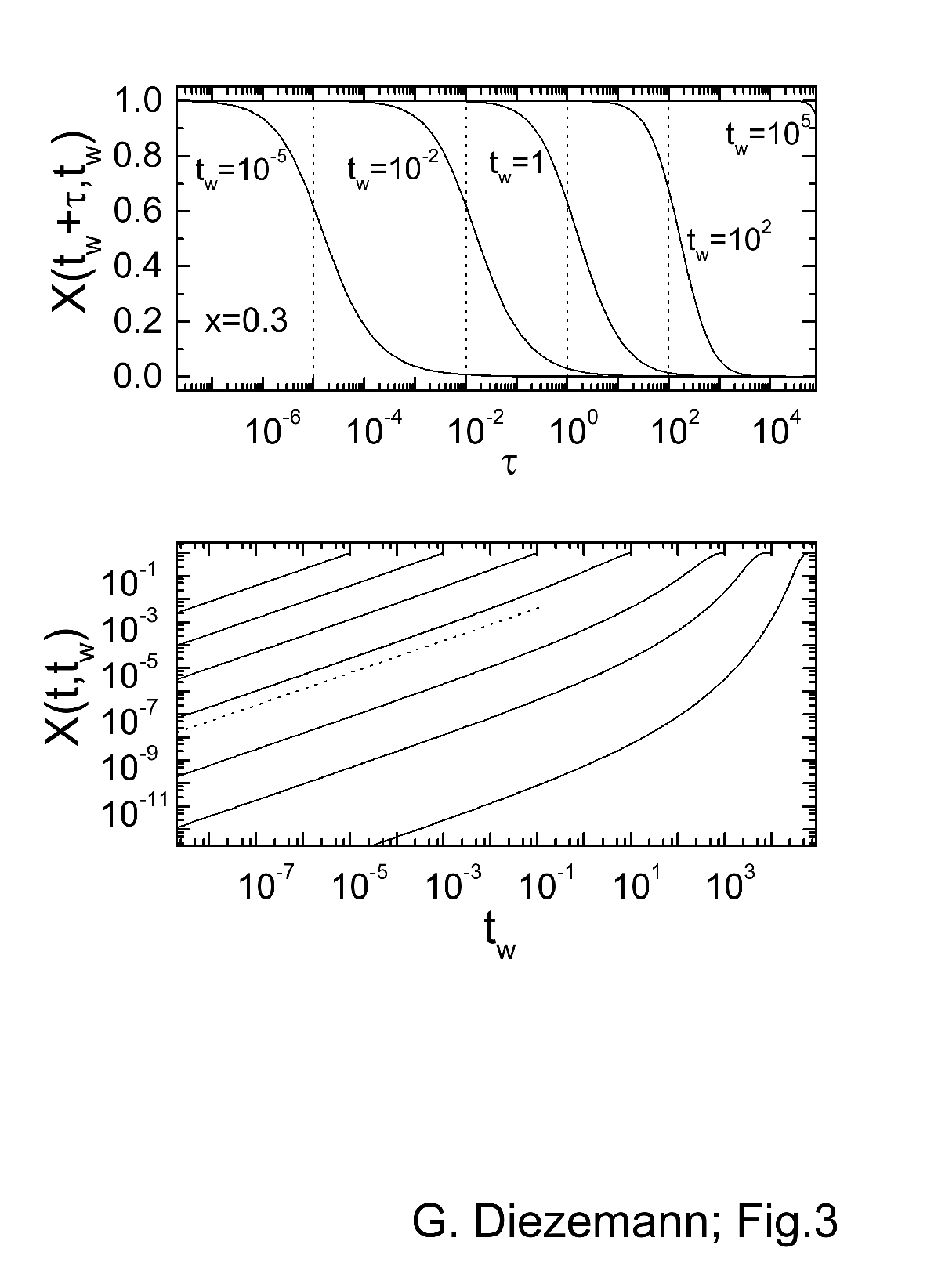}
\end{figure}
\newpage
\begin{figure}
\includegraphics[width=15cm]{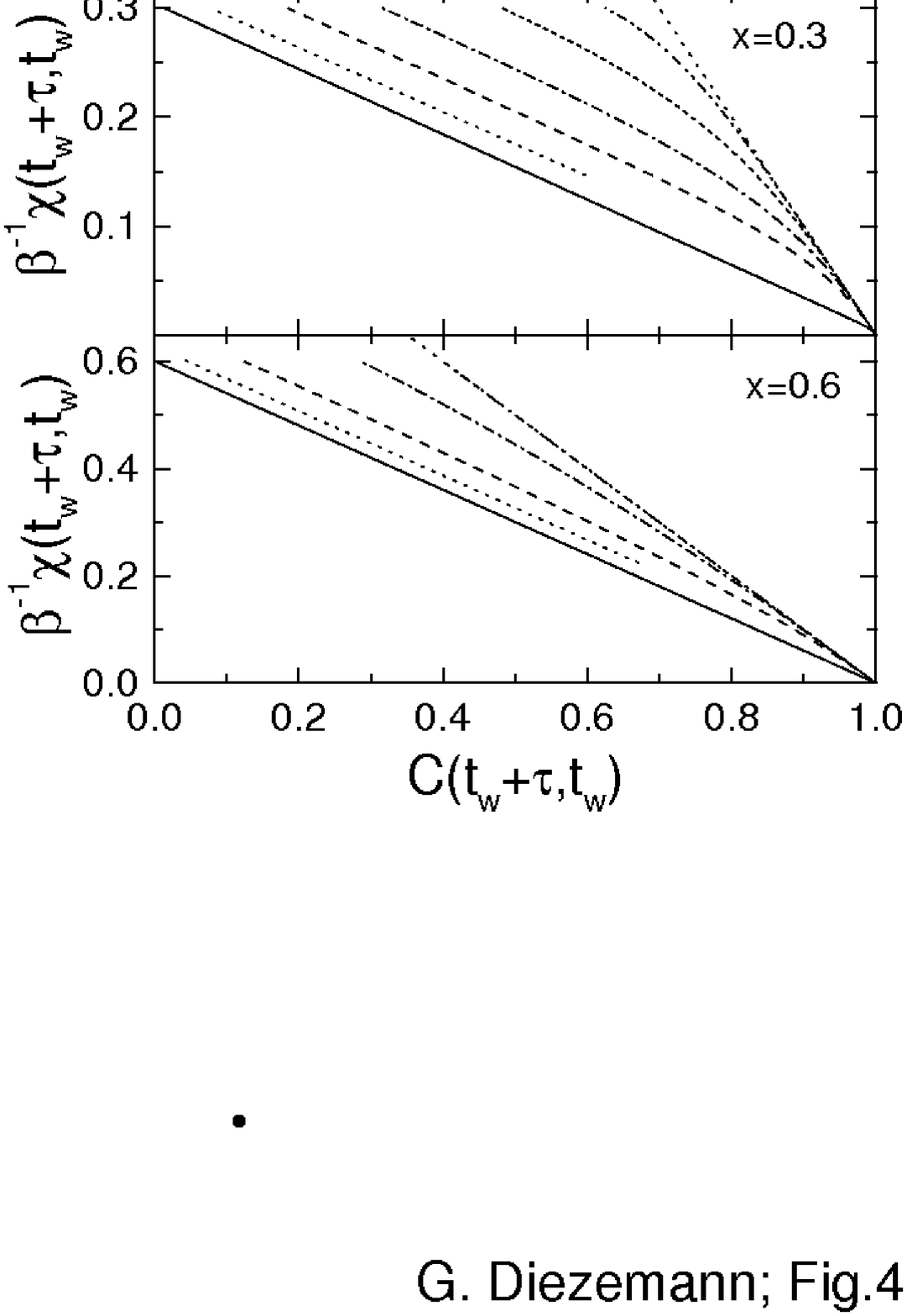}
\end{figure}
\end{document}